\documentclass[prl,twocolumn,showpacs,superscriptaddress,floatfix]{revtex4}
\usepackage{graphicx}

\usepackage{amsmath}
\usepackage{amssymb}
\usepackage{amsbsy}
\usepackage{textcomp}

\usepackage{enumerate}
\usepackage{color}
\definecolor{orange}{rgb}{1,0.5,0}
\renewcommand{\vec}[1]{\mathbf{#1}}
\newcommand{\gvec}[1]{\boldsymbol{#1}}
\newcommand{\vor}{\gvec{\omega}}
\newcommand{\vel}{\vec{u}}
\newcommand{\x}{\vec{x}}
\newcommand{\y}{\vec{y}}
\newcommand{\X}{\vec{X}}
\newcommand{\Alpha}{\gvec{\alpha}}
\newcommand{\vxi}{\gvec{\xi}}

\begin{document}

\title{Finite-Time Euler singularities: A Lagrangian perspective}

\author{Tobias \surname{Grafke}}
\affiliation{Theoretische Physik I, Ruhr-Universit\"at Bochum,
Universit\"atsstr. 150, D44780 Bochum (Germany)}
%\author{J\"urgen \surname{Dreher}}
%\affiliation{Theoretische Physik I, Ruhr-Universit\"at Bochum,
%Universit\"atsstr. 150, D44780 Bochum (Germany)}
\author{Rainer \surname{Grauer}}
\affiliation{Theoretische Physik I, Ruhr-Universit\"at Bochum,
Universit\"atsstr. 150, D44780 Bochum (Germany)}

\date{\today}

\begin{abstract}
  We address the question whether a singularity in a three-dimensional
  incompressible inviscid fluid flow can occur in finite
  time. Analytical considerations and numerical simulations suggest
  high-symmetry flows being a promising candidate for a finite-time
  blowup. Utilizing Lagrangian and geometric non-blowup criteria, we
  present numerical evidence against the formation of a finite-time
  singularity for the high-symmetry vortex dodecapole initial
  condition. We use data obtained from high resolution adaptively
  refined numerical simulations and inject Lagrangian tracer particles
  to monitor geometric properties of vortex line segments. We then
  verify the assumptions made by analytical non-blowup criteria
  introduced by Deng et. al [Commun. PDE \textbf{31} (2006)]
  connecting vortex line geometry (curvature, spreading) to velocity
  increase to rule out singular behavior.
\end{abstract}
\pacs{\textbf{47.10.-g}, \textit{47.10.ad}, 47.11.-j}
%\msc Primary 76B03 \sep Secondary 35Q31 \sep 65M06 \sep 65M55

%\pacs{47.27.Ak, 47.27.E-, 47.27.ef, 05.40.-a}

\maketitle

\paragraph{Introduction}

The incompressible Euler equations in three dimensions are
\begin{equation}
  \label{eq:euler}
  \frac{\partial \vel}{\partial t} + \vel \cdot \nabla \vel + \nabla p = 0\,, \quad \nabla \cdot \vel = 0\,.
\end{equation}
Existence and uniqueness of its solutions for all times are
unknown. Together with its prominent brother, the incompressible
Navier-Stokes equations, these equations have withstood the minds of
mathematicians and physicists for centuries. While the latter are
regarded as ``Millennium Prize Problem'' by the Clay Mathematics
Institute \citep{fefferman:2000}, the ignorance regarding existence of
global solutions is even larger for the inviscid case: The notion of
weak solutions, which is well established for the Navier-Stokes
equations since \citet{leray:1934}, is unknown for the
three-dimensional Euler equations.

As a now classical result, the blowup criterion of
\citet{beale-kato-majda:1984} (BKM) connects the existence of
solutions for the incompressible Euler equations in three dimensions
to the critical accumulation of vorticity. It has been tried in the
past to construct explicit initial conditions to obtain numerical
evidence for or against a finite-time singularity via BKM, with
surprisingly inconsistent results \citep{kerr:1993,hou-li:2006}. The
major reason for this ambiguity is the critical dependence on
extrapolation, which renders the identification of singular versus
near-singular behavior next to impossible by numerical means. The
hopes are high that the situation is less vague when considering
geometric analysis of the flow \citep{constantin-fefferman-majda:1996,
  cordoba-fefferman:2001, deng-hou-yu:2005, deng-hou-yu:2006}. In this
letter, we present the application of such geometric criteria to
numerical data to sharpen the distinction between singular and
near-singular flow evolution.

The Letter is organized as follows: we first review the notion of
geometric non-blowup criteria and state the considered criteria and
their interpretation. We then briefly name the computational setup and
implementation details of our numerical scheme to obtain adaptively
refined data of up to $8192^3$ mesh points. Using Lagrangian tracers
and diagnostics for vortex line geometry, we analyze the simulation
data to conclude a non-blowup for the considered initial conditions. A
conclusion and outlook summarize the Letter.

\paragraph{Geometric non-blowup criteria.}

Historically, non-blowup criteria for the incompressible Euler
equations commonly focus on global features of the flow, such as norms
of the velocity or the vorticity fields. This comes at the
disadvantage of neglecting the structures and physical mechanisms of
the flow evolution. A strategy to overcome such shortcomings was
established by focusing more on geometrical properties and flow
structures, such as vortex tubes or vortex lines. Starting with the
works of \citet{constantin:1994},
\citet{constantin-fefferman-majda:1996} , and
\citet{cordoba-fefferman:2001}, some of these geometric criteria
(e.g. \citep{gibbon:2002,deng-hou-yu:2005,gibbon-holm-kerr-roulstone:2006})
have reached a phase where they allow direct verification of their
assumptions with the help of numerical simulations.

Common to geometric criteria is the notion of vortex lines, defined as
integral curves along the vorticity direction field $\vxi$. They are
transported with the flow, i.e. two points $\x$ and $\y$ on the same
vortex line $c(s)$ stay on the same vortex line indefinitely. As
simple consequence of the solenoidality of $\vor$ and the BKM theorem,
one gets:

\textbf{Deng-Hou-Yu theorem 1:} \textit{Let $\x(t)$
  be a family of points such that for some $c_0>0$ it holds
  $|\omega(\x(t),t)|>c_0\Omega(t)$. Assume that for all $t \in [0,T)$
    there is another point $\y(t)$ on the same vortex line as $\x(t)$,
    such that the direction of vorticity $\vxi(\x,t) =
    \omega(\x,t)/|\omega(\x,t)|$ along the vortex line $c(s)$ between
    $\x(t)$ and $\y(t)$ is well-defined. If we further assume that}
\begin{equation}
  \left| \;\int_{\x(t)}^{\y(t)} \left( \nabla \cdot \vxi \right) (c(s),t) \,\mathrm{d}s \;\right| \leq C
\end{equation}
\textit{for some absolute constant C, and}
\begin{equation}
  \int_0^T |\omega(\y(t),t)|\,\mathrm{d}t < \infty\;,
\end{equation}
\textit{then there will be no blowup up to time $T$.}

This criterion can readily be applied to numerical simulations. On the
other hand, the same theorem may be interpreted in a different way to
distinguish between a point-wise blowup and the blowup of a complete
vortex line segment, as sketched in Fig. \ref{fig:T1}:
\begin{figure}[t]
  \includegraphics[width=0.7\linewidth]{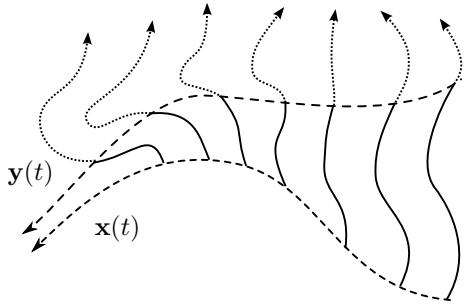}
  \caption{For the position $\x(t)$ of maximum vorticity, choose
    $\y(t)$ such that $\left| \;\int_{\x(t)}^{\y(t)} \left( \nabla
        \cdot \vxi \right) (c(s),t) \,\mathrm{d}s \;\right| = C$. For
    a point-wise singularity, $\x(t)$ and $\y(t)$ must collapse in finite
    time.\label{fig:T1}}
\end{figure}
At each instance in time, identify the point of maximum vorticity as
$\x(t)$. Now define $\y(t)$ such that $\left| \;\int_{\x(t)}^{\y(t)}
\left( \nabla \cdot \vxi \right) (c(s),t) \,\mathrm{d}s \;\right| = C$
for a constant threshold $C$. If a singularity occurs, then either
$\y(t)$ approaches $\x(t)$ (point-wise blowup), or, if the distance
between $\x(t)$ and $\y(t)$ stays finite, the complete vortex line
segment between $\x(t)$ and $\y(t)$ exhibits critical growth.

Results obtained with this method can be further improved by
considering the Lagrangian evolution of vortex line segments $L_t$ in
time. The geometric equivalent of the vortex stretching term is the
increase in length for a Lagrangian vortex line segment. It is
possible to quantify this stretching and establish a sound connection
to the vorticity dynamics of the flow.

\begin{figure}[t]
  \includegraphics[width=0.7\linewidth]{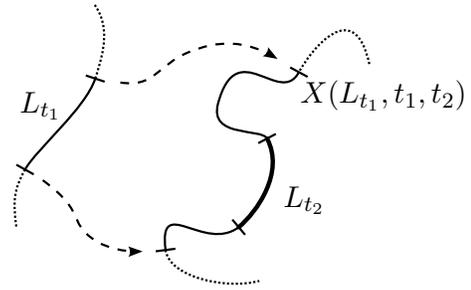}
  \caption{Lagrangian evolution of a vortex line segment $L_t$ in the
    context of theorem 2. For every $t_2 > t_1$, choose $L_{t_2}$ such
    that it is a subset of $\X(L_{t_1}, t_1, t_2)$.\label{fig:T2}}
\end{figure}

Denote with $l(t)$ the length of a vortex line segment $L_t$ at time
$t$ and define with $\Omega_L(t) := \|\vor(\cdot,t)\|_{L^\infty(L_t)}$
the maximum vorticity on the vortex line segment. Furthermore, let
$M(t) := \max(\|\nabla \cdot
\vxi\|_{L^\infty(L_t)},\|\kappa\|_{L^\infty(l_t)})$ be the quantity of
vortex line convergence $\nabla \cdot \vxi$ and vortex line curvature
$\kappa$, and define $\lambda(L_t) := M(t)l(t)$. It then holds (compare
Fig. \ref{fig:T2}):

\textbf{Deng-Hou-Yu theorem 2:} \textit{Assume there is a family of
  vortex line segments $L_t$ and $T_0 \in [0,T)$, such that $L_{t_2}
  \subseteq \X(L_{t_1}, t_1, t_2)$ for all $T_0 < t_1 < t_2 < T$. We
  also assume that $\Omega(t)$ is monotonically increasing and
  $\|\omega(t)\|_{L^\infty(L_t)} \geq c_0 \Omega(t)$ for some $c_0 >
  0$ when $t$ is sufficiently close to $T$. Furthermore, we assume
  that}
\begin{enumerate}[(i)]
\item $U_{\xi}(t) + U_n(t) \lambda(L_t) \lesssim (T-t)^{-A}$ \textit{for} $A \in (0,1)$
  \label{itm:T2_1}
\item $\lambda(L_t) \leq C_0,$
  \label{itm:T2_2}
\item $l(t) \gtrsim (T-t)^B$ \textit{for some} $B < 1-A$.
  \label{itm:T2_3}
\end{enumerate}
\textit{Then there will be no blowup in the 3D incompressible Euler
  flow up to time T.}

Here, $a(t) \lesssim b(t)$ means that there exists a constant $c \in
\mathbb{R}$ such that $|a(t)|~<~c\,|b(t)|$. The velocity components
are defined as $U_\xi(t) := \max_{\x, \vec{y} \in L_t} |(\vel \cdot
\vxi)(\x,t) - (\vel \cdot \vxi)(\vec{y},t)|$ and $U_n(t) :=
\max_{L_t}|\vel\cdot \vec{n}|$. The proof is given in
\citep{deng-hou-yu:2005}.

\begin{figure}[b]
  \centering
  \includegraphics[width=0.3\linewidth]{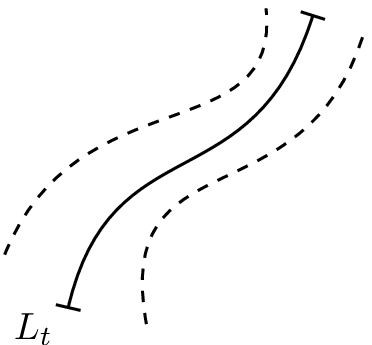}
  \includegraphics[width=0.3\linewidth]{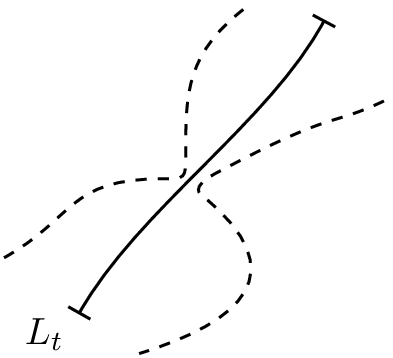}
  \includegraphics[width=0.3\linewidth]{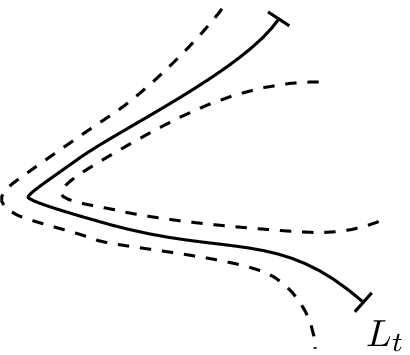}
  \caption[Characterizing vortex line geometry by
    $\lambda(L_t)$]{Characterizing vortex line geometry in terms of
    $\lambda(L_t)$. A slightly curved vortex line with approximately
    parallel neighboring vortex lines (\textbf{left}) exhibits small
    $\lambda(L_t)$. Vortex lines with tightening neighboring vortex
    lines (\textbf{center}) or vortex lines with high curvature, in
    comparison to their length (\textbf{right}) have large
    $\lambda(L_t)$.}
  \label{fig:T2lambda}
\end{figure}
All information about the geometric properties of the vortex line
segment under consideration is encoded in $\lambda(L_t)$,
characterizing the geometric ``tameness'' of the vortex line
filament. As depicted in Fig. \ref{fig:T2lambda}, a vortex line
segment has a huge $\lambda(L_t)$, if its maximum curvature is large,
relative to its length (the segment is ``kinked'' instead of
``curved''), or if the surrounding vortex lines collapse to the
considered segment in at least a point (the surrounding is
``tightening'' instead of ``parallel''). This quantifies the
constricted notion of ``relatively straight'' and ``smoothly
directed'' given in \citep{constantin-fefferman-majda:1996} in a
sharper way. The process of keeping $\lambda(L_t)$ bounded by some
constant $C_0$ translates to words as the process of ``zooming in'' to
the location of maximum vorticity in order to keep the considered
vortex line segment relatively straight in comparison to its
length. The assumed accompanying collapse in length to keep
$\lambda(L_t)$ bounded is then linked in its growth rate to the blowup
of the velocity components.

Despite high hopes from an analytical point of view that these
considerations will shed light on the true nature of vorticity
accumulation, numerical results observing geometrical properties of
Lagrangian vortex filaments are scarce. This is primarily due to the
fact that Eulerian quantities such as $\Omega(t)$ are readily
trackable in post-processing, while monitoring the Lagrangian
evolution requires additional computational effort. On top of that,
the geometry of integral curves at an instance in time, though in
principle computable in post-processing, as well as derived quantities
such as their convergence and curvature, are quite inaccessible in
comparison to simple Eulerian criteria.

\paragraph{Numerical Experiment}

\begin{figure}[t]
  \centering
  \includegraphics[width=0.42\linewidth]{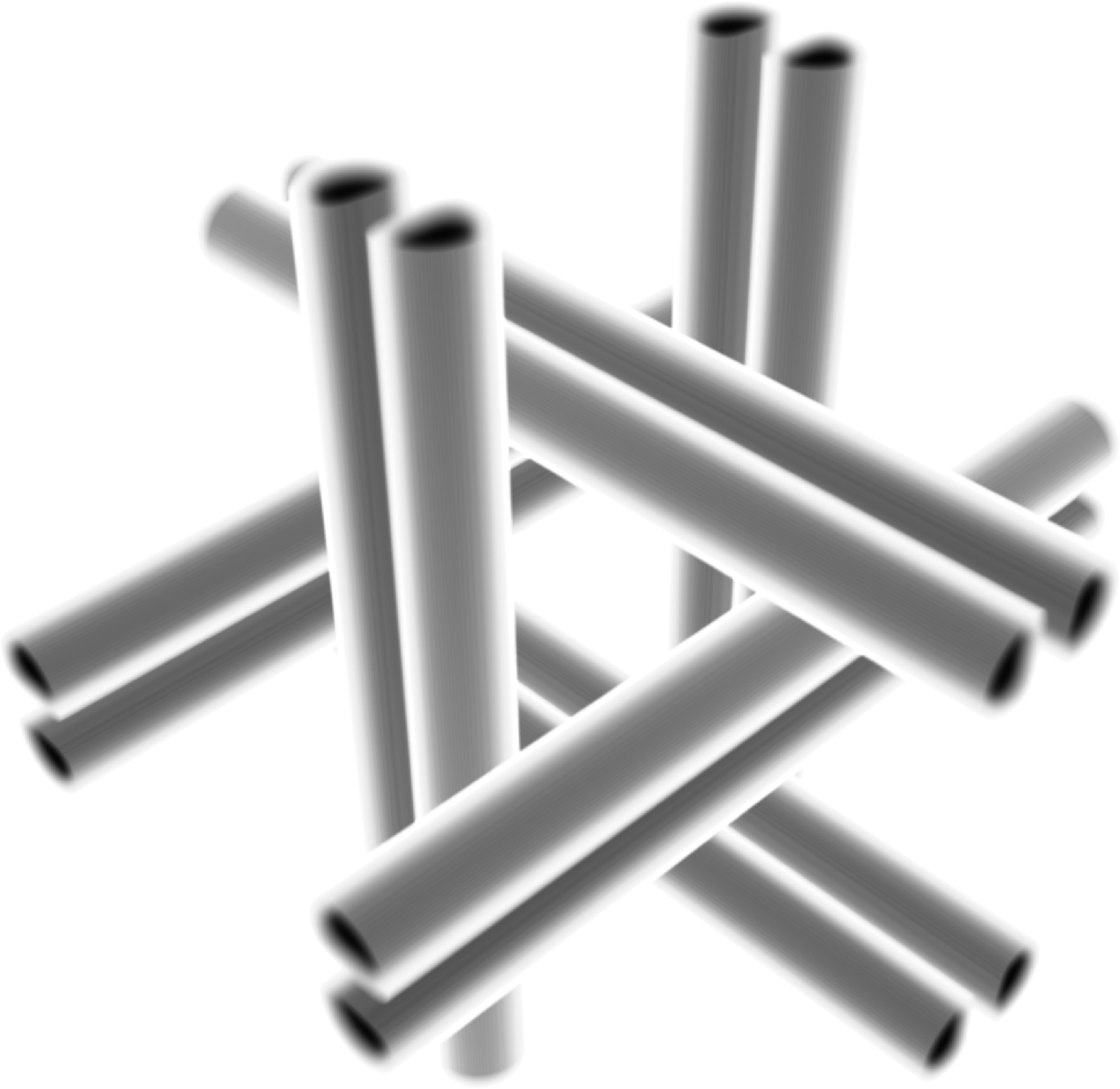}
  \hfill
  \includegraphics[width=0.55\linewidth]{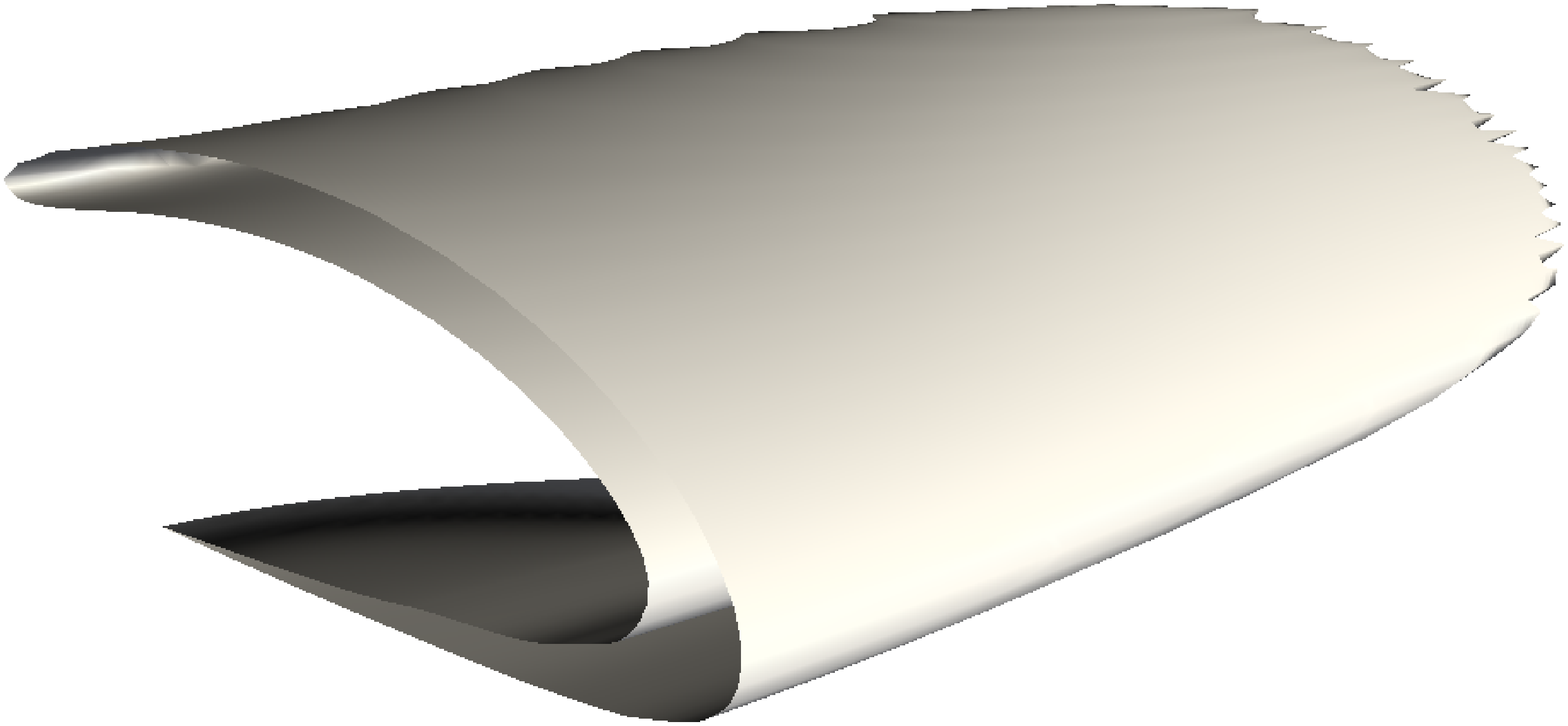}
  \caption{Left: Volume plot of the vorticity for the vortex
    dodecapole initial conditions: Six pairs of anti-parallel vortex
    tubes. Right: Isosurface of the vorticity for a single tube at a
    late time.\label{fig:initial}}
\end{figure}

Vorticity-strain coupling is the favored mechanism for the formation
of a finite-time Euler singularity. Since it is well established that
this process is inherently unstable for turbulent flows, it seems
natural to search for techniques to artificially keep the coupling
existent. One such technique is the introduction of symmetries to the
flow. Early examples such as the Taylor-Green vortex
\citep{taylor-green:1937} or Kerr's initial conditions
\citep{kerr:1993, hou-li:2006} are employing such symmetries. Yet, as
pointed out by \citet{pelz:2001}, for a single vortex tube to exhibit
critical vorticity-strain coupling, its curvature in the symmetry
plane has to blow up alongside the axial strain. On the other hand,
increasing axial strain diminishes the curvature of the critical
vortex line. These counteracting processes constitute an intrinsic
resistance of a single vortex line to locally ``self-stretch'' in a
critical way. The same argument holds for a pair of anti-parallel
vortex tubes. A popular way to counter this effect is to induce the
axial strain by neighboring tubes instead of relying on a sufficiently
large kink. This was accomplished, as suggested in
\citep{boratav-pelz:1994}, by introducing additional rotational
symmetries to the flow, arriving at the vortex dodecapole initial
condition pictured in Fig. \ref{fig:initial} (left). These initial
conditions are recognized as a promising candidate for the formation
of a finite-time singularity. As additional benefit, the high symmetry
introduces huge savings in computational effort and memory
requirements. An isosurface of the vorticity of one of the vortex
tubes at a late time is depicted in \ref{fig:initial} (right), showing
the typical roll-up of the vortex sheet in the critical region.

% state vorticity profile? (analytic form)
Resolution is paramount for a reliable statement on possible singular
behavior of the Euler equations. We use adaptively refined meshes
provided by the framework \textit{racoon} \citep{dreher-grauer:2005}
to reach effective resolutions of up to $8192^3$ mesh points,
scaling close to optimal on up to $10^5$ cores on massively parallel
machines. The numerical scheme consists of a strong stability
preserving third order Runge-Kutta \citep{shu-osher:1988} time
integrator combined with a third order shock-capturing CWENO scheme
\cite{kurganov-levy:2000} to reduce oscillations in the presence of
strong gradients. The integrated equation is the vorticity formulation
of the Euler equations,
%\begin{equation}
%  \label{eq:vorticity}
%  \frac{\partial}{\partial t} \vor + \nabla \times \left( \nabla (\vel \otimes \vel) \right) = 0\;,
%\end{equation}
employing a vector potential formulation $\Delta \vec{A} = -\vor$ with
$\vel = \nabla \times \vec{A}$ to ensure solenoidality of the
vorticity vector field $\vor$. The associated Poisson equation is
solved with a second order parallel and adaptive multigrid
algorithm. Interpolation on the coarse-fine interfaces is done in
$\vor$ to ensure the highest possible accuracy when applying the
aforementioned blowup-criteria. Passive tracer particles are injected
into the flow for the tracking of Lagrangian vortex line segments. The
above third order Runge-Kutta is also used for the time integration of
the tracer particles and the space integration of vortex
lines. Details of the numerical scheme, regarding its implementation,
adaptivity, parallelization, and diagnostics will be presented
elsewhere.

The numerical experiment now consists of two parts: First, utilizing
the method outlined above, we analyze the nature of the possible
singularity in terms of locality (point-wise versus filament). Then,
by means of the second theorem, we conclude from the scaling of length
and velocity components of the critical Lagrangian vortex filament, if
the observed behavior is singular at all.

\paragraph{Results}

For the first part, the constant $C$ of theorem 1 is chosen in a
reasonable way to achieve a length of the vortex line segment that
fits into the computational domain in the beginning of the simulation,
but is still well resolved at the chosen resolution at later
times. Hence, the whole vortex line segment is resolved reliably
throughout the simulation.

\begin{figure}[t]
  \centering
  \includegraphics[width=0.95\linewidth]{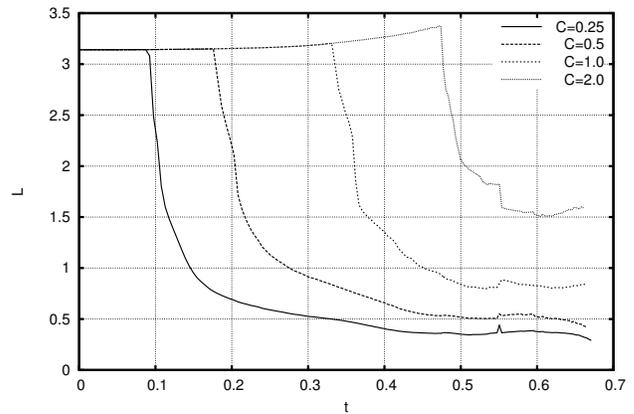}
  \caption{Length of the vortex lines starting at position $\x$ of
    maximum vorticity for different constants $C=\int_x^y \nabla \cdot
    \xi \mathrm{d}s$.\label{fig:T1length}}
\end{figure}
\begin{figure}[t]
  \centering
  \includegraphics[width=0.95\linewidth]{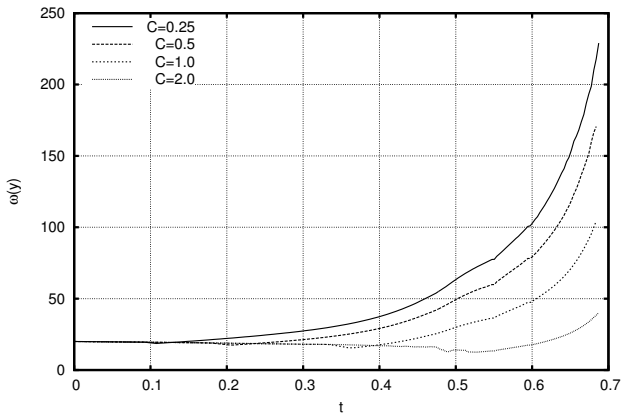}
  \caption{Vorticity at the endpoint $y$ of the considered vortex line
    segments. Once satiated, the growth rate is the same for all
    $y$. \label{fig:T1omega}}
\end{figure}

The results for the vortex dodecapole are presented in figure
\ref{fig:T1length} for different constants $C \in \{0.25, 0.5, 1,
2\}$. Initially, the vortex line segments do not accumulate enough
$\nabla \cdot \vxi$, so that the length is bounded by the size of the
computational domain ($\x \in [0,\pi]^3$). At some point, depending on
the value of $C$, the threshold is reached and the length of the
vortex line segment decreases. Yet, for all considered cases of $C$,
the length does not collapse to a point, but saturates at early times
without approaching $l(t) = 0$. This behavior appears to be stable up
to the latest time of the simulation. The final length of the vortex
line segments is at least $l=0.3$ for the smallest case of $C$
($C=0.25$), which is still well resolved with at least $200\,\Delta x$
for the simulation with $4096^3$ grid points. This result, therefore,
is a numerical evidence against a point-wise blowup for the vortex
dodecapole class of initial conditions. This is in concordance with
the estimate in \citep{deng-hou-yu:2005}. 

Yet, monitoring the development of $\vor(\y(t),t)$, as shown in figure
\ref{fig:T1omega}, yields a similar growth rate for the accumulation
of vorticity at the endpoint $\y(t)$ as for the beginning of the
vortex line segment $\x(t)$. This is hardly surprising, since by
construction a constant value for $C$ directly links the growth rates
of $|\vor(\x(t),t)|$ to $|\vor(\y(t),t)|$. Nevertheless, a numerical
verification of this analytic equality may be seen as a confirmation
that the observed growth rate of $|\vor(\x(t),t)|$ is by no means a
numerical artifact in an isolated small area, but is reproduced at
points far away from the critical region, which appear to be
well-behaved at first view. The possibly critical growth in the
perspective of BKM is, thus, confirmed by the global flow.

Furthermore, since for a large portion of the simulation the distance
$l(t)$ is approximately constant, this could be interpreted as an
evidence for the existence of a non-vanishing vortex line segment that
blows up in every point. Thus, contradicting the estimation of
\citep{deng-hou-yu:2005}, the possibility of a blowup of the vortex
dodecapole flow is not excluded by theorem 1. The scenario of a
collapse to a single point, on the other hand, is clearly conflicting
the numerical evidence.

For the second part of the numerical experiment, the above mentioned
theorem 2 is verified numerically to rule out a singularity in finite
time. The strategy is as follows:
\begin{itemize}
\item Identify the Lagrangian fluid element $\Alpha$, which will
  contain the maximum of vorticity at the latest time of the
  simulation, $\Omega(t) \approx |\vor(\X(\Alpha,t),t)|$. Numerically
  this procedure is implemented by carrying out a precursory identical
  simulation with a huge number of tracer particles ($\approx$ 1
  million) randomly distributed across the domain. Particles that
  accumulate huge amounts of vorticity are selected for the production
  run.
\item In a subsequent computation, at each instance in time, start a
  vortex line integration at $\X(\Alpha,t)$ along the vorticity
  direction field. Monitor the maximum curvature
  $\|\kappa\|_{L^\infty(L_t)}$ and the maximum vortex line convergence
  $\|\nabla \cdot \vxi\|_{L^\infty(L_t)}$ during the integration and
  calculate $\lambda(t)$. Stop the integration, as soon as
  $\lambda(t)$ reaches a fixed, arbitrary constant $C$. This defines
  $L_t$.
\item For this vortex line segment $L_t$, calculate the length $l(t)$,
  and the velocity components $U_n$ and $U_\xi$. From the collapse of
  the length $l(t)$ approximate the exponent $B$. This in turn
  provides the critical growth exponent $A$ for the velocity
  variables, $A_{\text{crit}} = 1 - B$.
\item Compare the increase in $U_n$ and $U_\xi$ to
  $1/(T-t)^{A_{\text{crit}}}$ to distinguish between critical and
  sub-critical growth of velocity.
\end{itemize}

This can be interpreted rather intuitively. By prescribing an
arbitrarily fixed $\lambda(t)$, the vortex line segment is kept
relatively geometrically uncritical, as the length-scale is always
adjusted accordingly. This process of ``zooming in'' just enough to
retain the geometric ``criticalness'' prescribes the rate of collapse
to a point, at least in the direction of the vortex line. All that is
left to check is whether the velocity growth in the immediate
surrounding is fast enough to be compatible with a finite-time
singularity.

\begin{figure}[t]
  \centering
  \includegraphics[width=\linewidth]{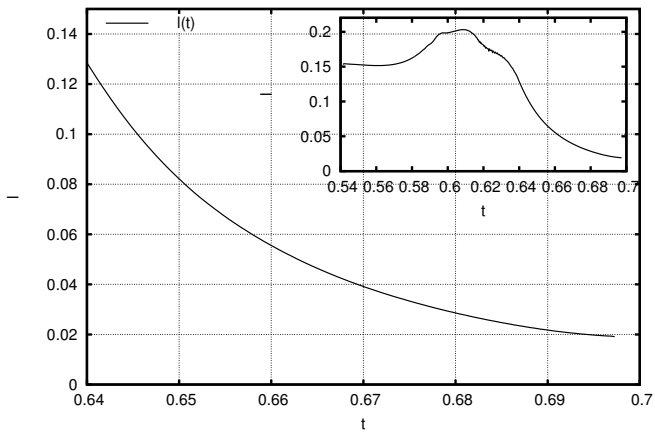}
  \caption[Evolution of the length of $L_t$]{Evolution of the length
    $l(t)$ of the critical vortex filament $L_t$ for different
    Lagrangian fluid elements. The length does not decrease as
    $(T-t)^B$ for any $B < 1$, which would be faster than linear. The
    Lagrangian collapse of the vortex segment is decelerating
    instead.}
  \label{fig:T2length}
\end{figure}
Figure \ref{fig:T2length} shows the results for the vortex dodecapole
initial conditions. Pictured is the length of the vortex line segment
for the tracer that is arriving at a position of very huge vorticity
at late stages of the simulation. The subplot depicts the long-term
behavior of the particle entering the critical region, while the final
stage of length decrease is magnified. The decrease in length does not
agree with a collapse in final time, but instead the shrinkage of the
segment decelerates clearly in time. This contradicts a scaling in
time proportional to $(T-t)^B$ for any $0 < B \le 1$, which would be
faster than (or, in the limiting case, equal to) linear.

\begin{figure}[t]
  \centering
  \includegraphics[width=\linewidth]{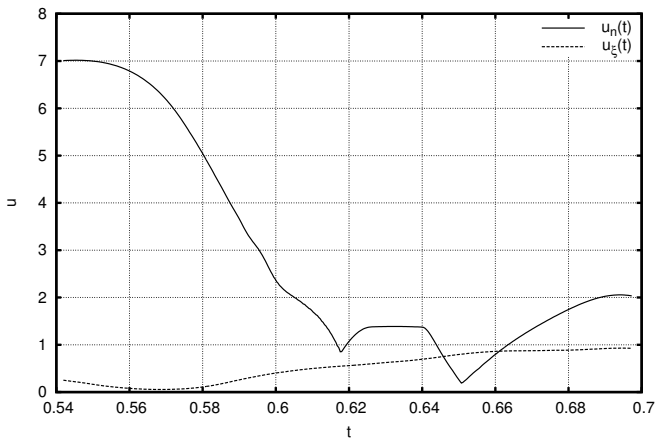}
  \caption{Evolution of the quantities $U_n$ and $U_\xi$ in time. Both
    quantities do not exhibit a critical growth of $1/(T-t)$.}
  \label{fig:T2u}
\end{figure}
It could furthermore be argued that the limit $B \rightarrow 0$ is
hard to exclude, since the drop in length would be virtually
instantaneous in time, with a close to constant scaling before. In
this limit, the quantities $U_n$ and $U_\xi$ would have to grow
roughly as $1/(T-t)$ to still allow formation of a finite-time
singularity. Figure \ref{fig:T2u} shows the observed behavior of $U_n$
and $U_\xi$ in time. Both show no signs of critical accumulation, in
particular not like $1/(T-t)$ in time. Thus, the assumptions of
theorem 2 are well met. These results therefore pose a strong evidence
against a finite-time singularity for the class of vortex dodecapole
initial conditions.

\paragraph{Conclusions and Outlook}

In this Letter we studied the question whether a singularity in a
three-dimensional incompressible inviscid fluid flow can occur in
finite time. Using massively parallel high-resolution adaptive
simulations and applying Lagrangian and geometric diagnostics to the
flow evolution, we are able to probe the behavior of the critical
vortex filament. Our findings pose a strong numerical evidence against
a finite-time singularity for the vortex dodecapole initial
conditions, based on analytical criteria connecting velocity scaling
to vortex line segment geometry. In principle, the presented method
could easily be applied to different classes of initial conditions.

\noindent {\bf Acknowledgment}
We would like to thank J. Dreher for his work on the computational
framework.  This work benefited from support through project
\mbox{GR~967/3-1} of the Deutsche Forschungsgesellschaft. Access to
the BlueGene/P multiprocessor computer JUGENE at the Forschungszentrum
J\"ulich was made available through project hbo35.

%\bibliography{bib}

\end{document}